\RequirePackage{fix-cm}
\documentclass[smallextended]{svjour3}       
\smartqed  
\usepackage{graphicx}

\begin{document}

\title{\rightline{\small IIT-CAPP-11-04}\\[0.05in]
Prospects for Antiproton Experiments at Fermilab\thanks{To appear in Proc.\ LEAP 2011, TRIUMF, Vancouver, BC, Canada,  April 27--May 1, 2011. 
}
}


\author{Daniel M. Kaplan       \\ 
        for the TAPAS and AGE Collaborations 
}


\institute{D.M. Kaplan \at
              Illinois Institute of Technology \\
              Tel.: +1 312 567-3389\\
              Fax: +1 312 567-3289\\
              \email{kaplan@iit.edu}           
}

\date{Received: date / Accepted: date}

\maketitle

\begin{abstract}
Fermilab operates the world's most intense antiproton source. Newly proposed experiments can use those antiprotons either parasitically during Tevatron Collider running or after the end of the Tevatron Collider program. For example, the annihilation of 5 to 8\,GeV antiprotons is expected to yield world-leading sensitivities to hyperon rare decays and {\em CP} violation. It could also provide the world's most intense source of tagged $D^0$ mesons, and thus the best near-term opportunity to study charm mixing and, via {\em CP} violation, to search for new physics. Other measurements that could be made include properties of the $X(3872)$ and the charmonium system.  An experiment using a Penning trap and an atom interferometer could make the world's most precise measurement of the gravitational force on antimatter. These and other potential measurements using antiprotons offer a great opportunity for a broad and exciting physics program at Fermilab in the post-Tevatron era.
\end{abstract}

\section{Introduction}
\label{intro}
The Fermilab Antiproton Source now produces more than $1.5\times10^{15}$ antiprotons per year~\cite{Pasquinelli-etal}. As Table~\ref{tab:sources} indicates, this substantially exceeds the intensity  available at the CERN Antiproton Decelerator (AD), as well as that anticipated at the Facility for Antiproton and Ion Research (FAIR) in Darmstadt, Germany. After Tevatron running ends, an internal target could  be operated in the Fermilab Antiproton Accumulator (as was done previously for Experiments 760 and 835), allowing fixed-target experimentation with beam kinetic energy in the range $\approx\!3.5$--8\,GeV. With antiproton stacking 10--20\% of the time, a luminosity of $2\times10^{32}\,{\rm cm}^{-2}{\rm s}^{-1}$ could be sustained during the remaining $\approx$\,80\%. Such a program could allow world-leading studies of hyperon rare decays and {\em CP} violation, the $X(3872)$ and other ``mystery states" in the charmonium region, and the charmonium system. While the open-charm production cross section at these energies has not been measured,  world-leading studies of charm mixing, rare decays, and {\em CP} violation may also be possible. (For brevity, several additional physics topics are omitted here;  for more detailed discussions, see the ``New pbar Experiments for Fermilab" website~\cite{website}.)

\begin{table}
\caption{Properties of existing and anticipated antiproton sources}
\label{tab:sources}       
\begin{tabular}{lccccc}
\hline\noalign{\smallskip}
& ${\overline{p}}$ & \multicolumn{2}{c}{{Stacking:}} & \multicolumn{2}{c}{{Operation:}}  \\
$\!\!$Facility & Kinetic Energy& {Rate}  & Duty & {Hours} & ${\overline{p}}$/{yr}\\
 & (GeV) & $(10^{10}/$hr) &  Factor & {/yr} &  ${(10^{13})}$\\
\noalign{\smallskip}\hline\noalign{\smallskip}
& 0.005 \\
\raisebox{1.5ex}[0pt]{$\!\!$CERN AD}& 0.047 & \raisebox{1.5ex}[0pt]{--} & \raisebox{1.5ex}[0pt]{--} & \raisebox{1.5ex}[0pt]{3800} &\raisebox{1.5ex}[0pt]{0.4} \\
\multicolumn{3}{l}{$\!\!$Fermilab Accumulator:}\\
~current operation& 8 & $>$\,25 & 90\% & 5550 & $>$\,150\\
~proposed here &$\approx$\,3.5--8& 20 & 15\% & 5550 &17 \\
$\!\!$FAIR ($\stackrel{>}{_\sim}\,$2018*) & 1--14 &  3.5 & 15\%* & 2780* & 1.5\\
\noalign{\smallskip}\hline
\end{tabular}
\begin{quotation}\footnotesize\noindent
$^*$The number of operating hours at FAIR reflects sharing of the collection ring  between the antiproton and radioactive-beam programs. With the staged FAIR construction plan, antiproton stacking will occur in the experiment ring, leading to a small stacking duty factor, as indicated here, until the stacking ring is built.
\end{quotation}

\end{table}

\section{Hyperon {\em CP} violation and rare decays}

\begin{table}
\caption{Summary of predicted hyperon {\em CP} asymmetries}
\label{tab:HyCP}       
\begin{tabular}{lccccc}
\hline\noalign{\smallskip}
 Asymm. & Mode & SM & Ref. & NP & Ref. \\
\noalign{\smallskip}\hline\noalign{\smallskip}
$A_\Lambda$ & $\Lambda\to p\pi$ & $\stackrel{<}{_\sim}4\times10^{-5}$ &  \cite{Tandean-Valencia-PRD67} & $\stackrel{<}{_\sim}6\times10^{-4}$ & \cite{Chang-etal}\\
$A_{\Xi\Lambda}$ & $\Xi^\mp\to\Lambda\pi$, $\Lambda\to p\pi$ & $\stackrel{<}{_\sim}5\times10^{-5}$ & \cite{Tandean-Valencia-PRD67} & $\le1.9\times10^{-3}$ & \cite{He-etal-SUSY} \\
$A_{\Omega\Lambda}$ & $\Omega\to\Lambda K$, $\Lambda\to p\pi$ & $\le4\times10^{-5}$ & \cite{Tandean} & $\le8\times10^{-3}$ & \cite{Tandean}\\
$\Delta_{\Xi\pi}$ & $\Omega\to\Xi^0 \pi$ & $2\times10^{-5}$ &\cite{Tandean-Valencia}& $\le2\times10^{-4}\,^*$ &  \cite{Tandean-Valencia}\\
$\Delta_{\Lambda K}$ & $\Omega\to\Lambda K$ & $\le1\times10^{-5}$ &\cite{Tandean}&  $\le1\times10^{-3}$ & \cite{Tandean} \\
\noalign{\smallskip}\hline
\end{tabular}
\begin{quotation}\footnotesize\noindent $^*$Once they are taken into account,  large final-state interactions are expected to increase this prediction to a range comparable to that for $\Omega\to\Lambda K$~\protect\cite{Tandean-private}.
\end{quotation}
 \end{table}

\label{sec:1}
{\em CP} violation (CPV) in hyperon decay is expected at the $\stackrel{<}{_\sim}$\,10$^{-5}$ level in the Standard Model and can be  enhanced by one to two orders of magnitude in models with new physics (see Table~\ref{tab:HyCP}). Searches for hyperon CPV are complementary to studies of the $K^0$ and beauty systems; for example, hyperon and $K^0$ CPV probe new-physics phases in parity-conserving (violating) currents, respectively. With the HyperCP (Fermilab E871)~\cite{E871} result,  $A_{\Xi\Lambda}\approx(\alpha_\Xi\alpha_\Lambda-\alpha_{\overline \Xi}\alpha_{\overline \Lambda})/(\alpha_\Xi\alpha_\Lambda+\alpha_{\overline \Xi}\alpha_{\overline \Lambda})=(-6.0\pm2.1\pm2.1)\times10^{-4}$~\cite{Materniak}, the most sensitive to date, experimental sensitivities have reached the few\,$\times10^{-4}$ level in the $\Xi^\mp\to{}^{{}^(}{\overline \Lambda}{}^{{}^{\,)}}\!\pi^\mp\to {}^{{}^(}
{\overline p}{}^{{\,}^)}\!\pi^\mp\pi^\mp$ decay chain.
 
 HyperCP also observed for the first time the flavor-changing neutral-current decay $\Sigma^+\to p\mu^+\mu^-$~\cite{Parketal}. The three observed events displayed a surprisingly narrow dimuon-mass distribution, suggesting the possibility of a new 
 pseudoscalar or axial-vector resonance as an intermediate state: $\Sigma^+\to p P^0,$ $P^0\to \mu^+\mu^-$, with $P^0$ mass of $(214.3\pm0.5)$\,MeV/$c^2$~\cite{Parketal}. Such a state could not be an ordinary meson, but could arise in models with new physics~\cite{P0models}. Given the small number of observed events, the effect could alternatively be a $\approx$\,2.4$\sigma$ fluctuation of the Standard Model virtual-photon coupling.

These topics motivate an experiment with substantially higher hyperon statistics than HyperCP. This cannot be accomplished in any currently operating or approved experiment. It could be done with fixed-target running of the Antiproton Accumulator, whose beam can be decelerated to just above the ${\overline p}p\to\Omega^-{\overline\Omega}{}^+$ threshold (Table~\ref{tab:thresh}). A 1-year run at $2\times10^{32}\,{\rm cm}^{-2}{\rm s}^{-1}$ luminosity should produce some $10^8$ $\Omega^-{\overline\Omega}{}^+$ pairs, giving statistical sensitivities of $\approx  7.8\times10^{-5}$ and $1.3\times10^{-4}$, respectively, for the {\em CP}-violating observables,
\begin{eqnarray*}
\Delta_{\Lambda K}&\equiv&\frac{\Gamma(\Omega^-\to\Lambda K^-)-\Gamma({\overline \Omega}{}^+\to{\overline \Lambda} K^+)}{\Gamma(\Omega^-\to\Lambda K^-)+\Gamma({\overline \Omega}{}^+\to{\overline \Lambda} K^+)}~,\\
\Delta_{\Xi\pi}&\equiv&\frac{\Gamma(\Omega^-\to\Xi^0 \pi^-)-\Gamma({\overline \Omega}{}^+\to{\overline \Xi}{}^0 \pi^+)}{\Gamma(\Omega^-\to\Xi^0 \pi^-)+\Gamma({\overline \Omega}{}^+\to{\overline \Xi}{}^0 \pi^+)}\,.
\end{eqnarray*}
Estimates of the systematic uncertainties are still in progress, but it appears that the uniquely clean environment of ${\overline p}p$ annihilation just above threshold will permit measurements at the $10^{-4}$ level (cf.~\cite{SuperLEAR}).

\begin{table}[t]
\caption{Measured and estimated ${\overline p}p\to$\,hyperon-antihyperon cross sections just above threshold.}\label{tab:thresh}
\begin{tabular}{lccccc}
\hline\noalign{\smallskip}
& \multicolumn{3}{c}{~~$\sqrt{s}$,~~ {$\overline p$} momentum and K.E.} &  & \\
\raisebox{1.5ex}[0pt]{Reaction} & (GeV) &  (GeV/$c$) & (GeV) & \raisebox{1.5ex}[0pt]{Cross section} & \raisebox{1.5ex}[0pt]{Ref.} \\
\noalign{\smallskip}\hline\noalign{\smallskip}
${\overline p}p\to\Lambda{\overline \Lambda}$& 2.304 & ~1.642 &0.953& $\approx65$\,$\mu$b & \cite{Johansson} \\
${\overline p}p\to\Xi^-{\overline \Xi}{}^+$ & 2.77 & ~3.0 & 2.2 & $\approx2$\,$\mu$b$^*$ & \cite{Baltay,HERAG}\\
${\overline p}p\to\Omega^-{\overline \Omega}{}^+$ & 3.39 & ~5.1 &4.3& $\approx60$\,nb & \\
\noalign{\smallskip}\hline
\end{tabular}
\begin{quotation}
\footnotesize\noindent
$^*$While the cross section at 3.0\,GeV/$c$ $\overline{p}$ momentum has not been measured, that at 3.5\,GeV/$c$ has been and is shown here.
\end{quotation}
\end{table}

Given the above-mentioned $2\sigma$ indication of possible CPV in $\Xi^\mp\to\Lambda\pi\to p\pi\pi$ decay~\cite{Materniak}, it would also be desirable to decelerate antiprotons to just above the $\Xi^-{\overline \Xi}{}^+$ threshold. This should be possible in the Accumulator; the key question is with what efficiency. The E835 collaboration developed the ``snowplow" technique to retune the lattice while decelerating, in order to avoid transition-induced beam losses~\cite{McGinnis-Stancari-Werkema}. Whether the method can be extended so low in energy (Table~\ref{tab:thresh}) remains to be seen, based on tests that can only be performed once the Tevatron run finishes.

\section{Measurements in the charmonium region}

\begin{figure}
\includegraphics[width=.5\columnwidth
,trim=0 2 0 5,clip]{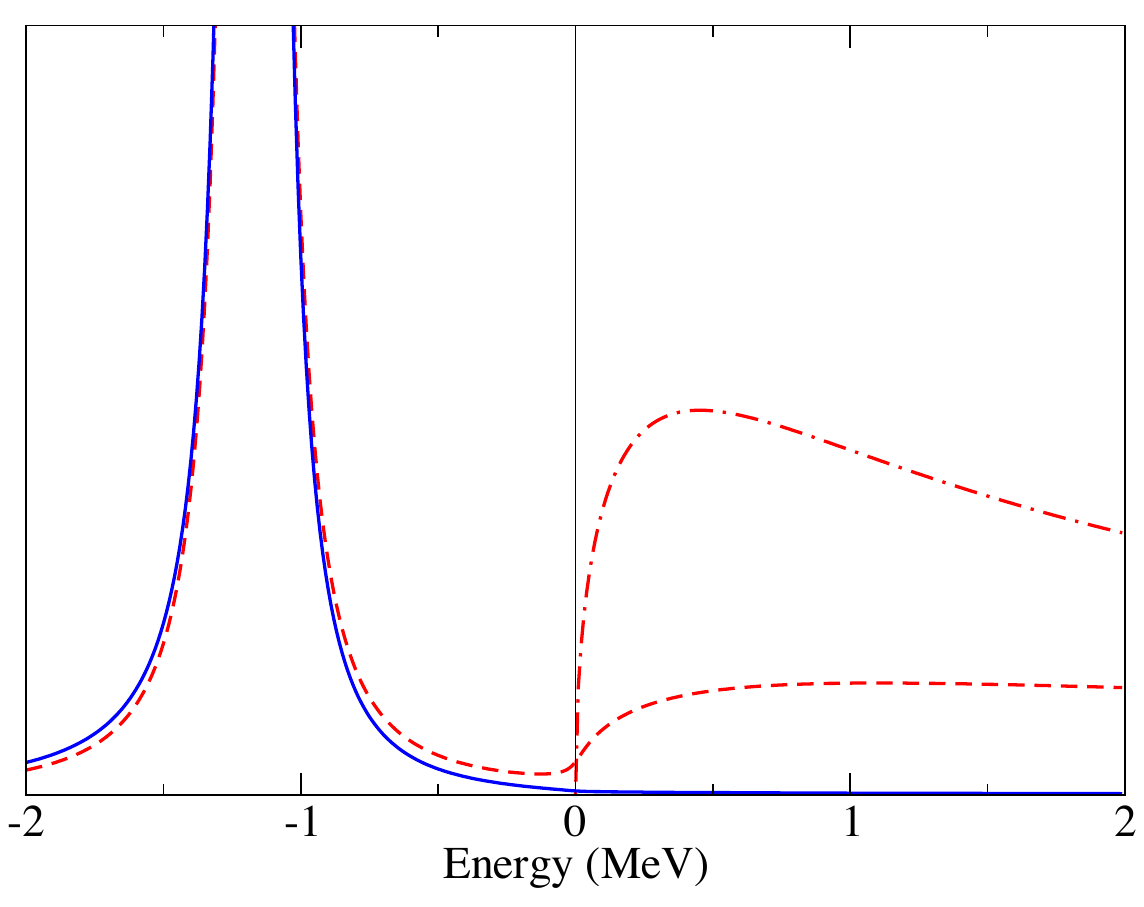}\includegraphics[width=.509\columnwidth
]{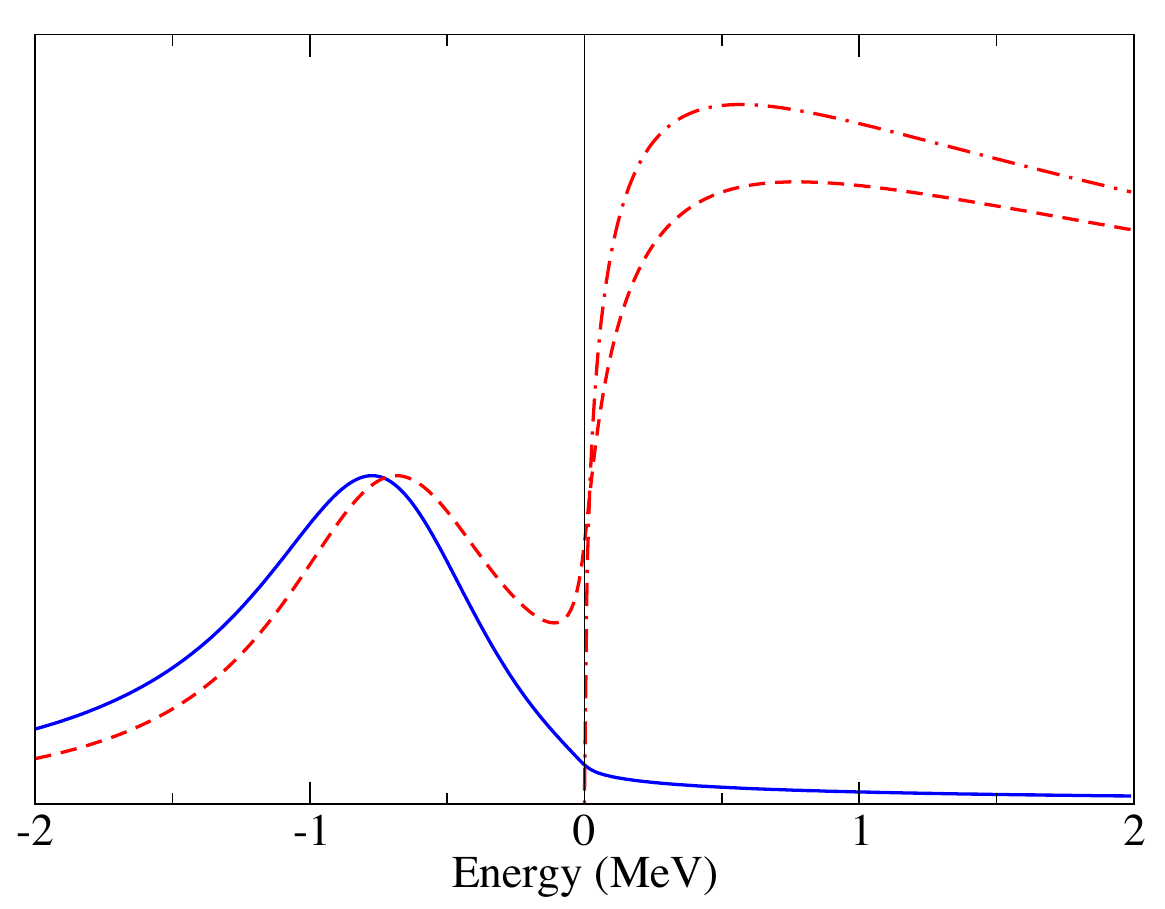}
\caption[Examples of expected $X(3872)$ lineshapes in $J/\psi\pi^+\pi^-$  and $D^0{\overline D}{}^0\pi^0$ final states in the molecular hypothesis]{Examples of expected $X(3872)$ lineshapes in $J/\psi\pi^+\pi^-$ (solid-blue curve) and $D^0{\overline D}{}^0\pi^0$ (dashed- and dot-dashed-red) final states for various parameter choices in the molecular hypothesis (from \protect\cite{Braaten-Stapleton}).}\label{fig:lineshape}
\end{figure}

Using the Fermilab Antiproton Source,  experiments E760 and E835 made the world's most precise measurements of charmonium masses and widths~\cite{E760-chi,E835-psi-prime}. This ($<$\,100\,keV) precision was enabled by the  small energy spread of the stochastically cooled antiproton beam and the absence of Fermi motion and negligible energy loss in the H$_2$ cluster-jet target. 
Although charmonium has by now been extensively studied, a number of questions remain in this region, most notably the nature of the mysterious $X(3872)$ state~\cite{ELQ} and improved measurement  of $h_c$ and $\eta^\prime_c$ parameters~\cite{QWG-Yellow}. The width of the $X$ may well be small compared to 1\,MeV~\cite{Braaten-Stapleton}. The unique precision of the ${\overline p}p$ energy-scan technique is ideally suited to making the precise mass, lineshape, and width measurements needed to test the intriguing hypothesis that the $X(3872)$ is a $D^{*0}\overline{D}{}^0$ molecule~\cite{molecule}. As shown in Fig.~\ref{fig:lineshape}, in the molecular hypothesis, the lineshape of the $X(3872)$ will be distinctive and will depend on decay mode. For optimal $\sqrt{s}$ resolution, these measurements will require the use of a hydrogen target: either an improved version of the E835 gas jet or a windowless, frozen-hydrogen target~\cite{Ishimoto-private,Ishimoto} (see below).

The formation cross section of $X(3872)$ in ${\overline p}p$ annihilation has not been measured, but it has been estimated to be similar in magnitude to that of the $\chi_c$ states~\cite{ditto,Braaten-X-3872}. By extrapolation from E760, about 500 
events would be observed in the $X(3872) \to\pi^+\pi^- J/\psi$ mode per day at the peak of the $X (3872)$. 
Given the uncertainties 
in the cross section and branching ratios, this may well be an under- or overestimate 
of the ${\overline p}p$ formation and observation rates, perhaps by as much as an order of magnitude. 
Nevertheless, it appears that a new experiment at the Antiproton Accumulator could obtain 
the world's largest clean samples of $X (3872)$, in perhaps as little as a month of running.\footnote{Although CDF and D\O\ are 
amassing samples of order $10^4$ $X (3872)$ decays, the large backgrounds in the CDF and D\O\ 
observations limit 
their incisiveness.}
In a few months of running, hundreds to thousands of observed events can be expected in all of the known decay modes, and many more, as-yet-unknown, modes should be seen as well. We will also have the opportunity to study the angular distributions of both the known and unknown modes. 
The high statistics, event cleanliness, and unique precision available in the ${\overline p}p$ formation 
technique could enable the world's smallest systematics. This experiment could thus 
provide a definitive test of the nature of the $X (3872)$.


\section{Charm mixing, {\em CP} violation, and rare decays}

\begin{figure}
\includegraphics[width=.5\linewidth]
{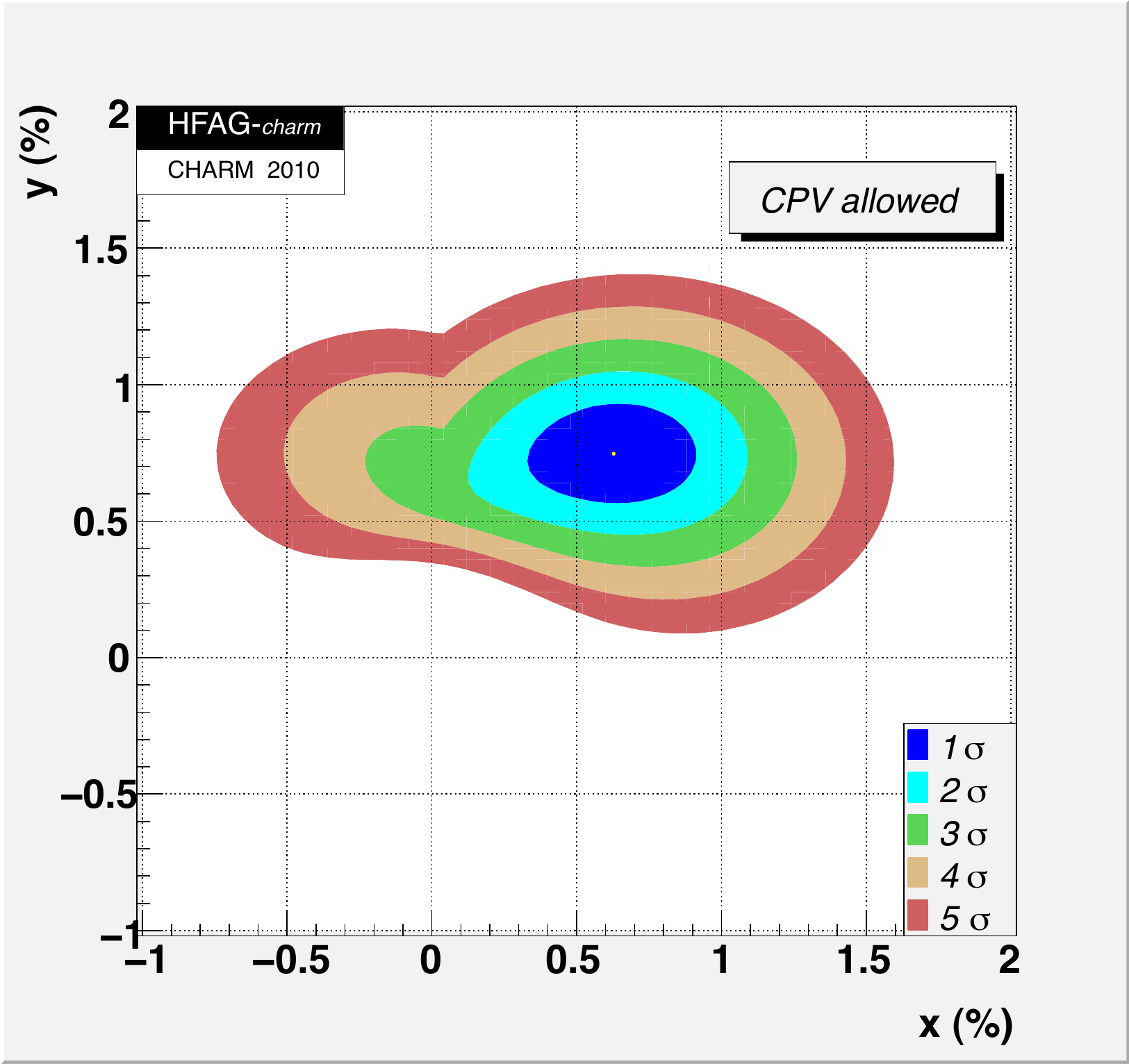}
\caption{World average of $D^0$--${\overline D}{}^0$ mixing parameters $x\equiv\Delta m/\Gamma$, $y\equiv\Delta\Gamma/2\Gamma$: best-fit values are $x=\left(0.63^{+0.19}_{-0.20}\right)$\%, $y=(0.85\pm0.12)$\%, and no mixing $(x=y=0)$ is disfavored by 10.2$\sigma$ (from~\protect\cite{HFAG}).
}\label{fig:mixing}
\end{figure}

After a $>$\,20-year search, $D^0$--${\overline D}{}^0$ mixing is now established  at $>$\,10 standard deviations (Fig.~\ref{fig:mixing})~\cite{HFAG,ICHEP2010}, thanks  to the $B$ factories and CDF. The level of mixing ($\sim$\,1\%) is consistent with the wide range of Standard Model predictions~\cite{Bigi-Uraltsev}; however, this does not preclude a significant and potentially detectable contribution from new physics~\cite{Bigi09,Grossman-etal}. Since some new-physics models predict differing effects in the charge-2/3 (``up-type") and --1/3 quark sectors~\cite{Bigi09,Grossman-etal}, it is important to carry out such  studies not only with $s$ and $b$ hadrons, but with charm mesons as well\,---\,the only up-type system for which  meson mixing can be measured.

While the total charm-production cross section for $\approx$\,8\,GeV antiprotons incident on proton or nucleon targets is challenging to compute from first principles, recent phenomenological estimates imply  values in the 1--10\,$\mu$b range~\cite{Braaten-X-3872,PANDA-LoI,Quigg-Eichten,Braaten-charm-est,Titov-Kampfer,Titov}. This is sufficiently large that the experiment we propose could amass a sample  ten or more times larger than  those of the B factories, years before the super-B factories reach comparable sensitivities.
For example, model calculations of the exclusive cross section $\sigma(\overline{p}p\to D^{*0}{\overline D}{}^0)$ peak at about $1\,\mu$b at $\sqrt{s} \approx 4.2$\,GeV~\cite{Braaten-charm-est,Titov-Kampfer,Titov}. This  corresponds to antiprotons of 8\,GeV kinetic energy (the Antiproton Source design energy) impinging on a fixed target and, at ${\cal L}=2\times10^{32}$\,cm$^{-2}$s$^{-1}$,  represents some $4\times10^9$ events produced per year. Since there will  also be $D^{*\pm}D^\mp$, $D^*\overline{D}{}^*$,  $D\overline{D}$,  $D\overline{D}\pi$,...\ events,  the total charm sample will be yet larger, and with the use of a target nucleus heavier than hydrogen, the charm-production $A$-dependence~\cite{A-dep,Lourenco} should enhance statistics by a further factor of a few. The total sample could thus substantially exceed the $10^9$ events
now available at the $B$ factories. 
Indeed, we project in Table~\ref{tab:D-prod}  in excess of $10^{10}$ tagged-$D^0$ events produced per year of running.

By localizing the primary interactions to  $\sim$\,10\,$\mu$m along the beam direction, a thin wire or frozen-hydrogen target can allow the $D$ decay to be resolved. The low charged multiplicity at these energies~\cite{PDG} implies small combinatorial background, so that  clean samples can be amassed using only modest vertex cuts, and thus, with high efficiency. Medium-energy 
$\overline{p}p$ or $\overline{p}N$ annihilation may thus be the optimal way to study charm mixing, and to search for possible new-physics contributions via the clean signature~\cite{Bigi09,Petrov,Buchalla-etal} of charm CPV.

\subsection[${D^0}$ mixing]{${ D^0}$ mixing}

Several signatures for $D^0$--${\overline D}{}^0$ mixing have been observed and indicate that it is at the upper end of the range expected in the SM~\cite{PDG}. These involve differing time-dependences of ``right-sign" 
Cabibbo-favored and ``wrong-sign" 
$D^0$ decays (arising  both from doubly Cabibbo-suppressed decay and from mixing), differing lifetimes of  decays to {\em CP}-even and mixed-$\!${\em CP} final states, and Dalitz-plot analyses of 3-body $D^0$ decays. 
These processes are sensitive to various combinations of the reduced mixing parameters $x\equiv\Delta m/\Gamma$, $y\equiv\Delta\Gamma/2\Gamma$. As already mentioned,  mixing at the observed level  could be due to SM physics, but there could also be an appreciable or even dominant contribution from new physics, which could be indicated by {\em CP} violation.

\begin{table}[t]
\caption[Example sensitivity estimate for $D^*$-tagged $D^0\to K\pi$ decays.]{Example sensitivity estimate for $D^*$-tagged $D^0\to K\pi$ decays (after Ref.~\cite{Braaten-X-3872}). 
}
\label{tab:D-prod}
\begin{tabular}{ccc}
\hline\noalign{\smallskip}
Quantity & Value & Unit \\ 
\noalign{\smallskip}\hline\noalign{\smallskip}
Running time & $2\times10^7$ & s/yr\\
Duty factor & 0.8* & \\
$\cal L$ & $2\times10^{32}$& cm$^{-2}$s$^{-1}$ \\
\noalign{\smallskip}\hline\noalign{\smallskip}
Annual integrated $\cal L$ & 3.2 & fb$^{-1}$ \\
\noalign{\smallskip}\hline\noalign{\smallskip}
Target $A$ & 47.9 & \\
$A^{0.29}$ & 3.1 & \\
$\sigma({\overline p}p\to D^{*+}+{\rm anything})$ & 1.25--4.5 & $\mu$b \\
\noalign{\smallskip}\hline\noalign{\smallskip}
\# $D^{*\pm}$ produced & (0.3--1)$\,\times 10^{11}$ & events/yr \\
\noalign{\smallskip}\hline\noalign{\smallskip}
${\cal B}(D^{*+}\to D^0\pi^+)$  &0.677 & \\
${\cal B}(D^{0}\to K^-\pi^+)$ & 0.0389 & \\
Acceptance & 0.45 & \\
Efficiency & 0.1--0.3 & \\
\noalign{\smallskip}\hline\noalign{\smallskip}
Total & (0.3--3)$\,\times10^8$ & events/yr\\
\noalign{\smallskip}\hline
\end{tabular}
\begin{quotation}
\footnotesize\noindent
{$^*$\footnotesize\noindent Assumes $\approx$\,15\% of running time is devoted to antiproton-beam stacking.}
\end{quotation}
\end{table}

Given the kinematic similarities between the B factory $D$ samples and that in our proposed experiment, we anticipate performing all of these mixing analyses with significantly greater sensitivity than has been achieved heretofore. Table~\ref{tab:D-prod} gives an example sensitivity calculation in $D^*$-tagged $D^0\to K\pi$ decays. Our sensitivity in semileptonic decays will depend on the efficiency and purity of lepton identification, which we have not yet simulated. In hadronic modes, we could be the world's most sensitive experiment, exceeding current B-factory statistics by a factor of 10 or more, and perhaps in semileptonic modes as well. Depending on their trigger and reconstruction efficiencies for charm, LHC$b$ may achieve statistical sensitivities comparable to or exceeding ours, but we expect them to have appreciable systematic uncertainties for small ($\stackrel{<}{_\sim}10^{-3})$ charm CPV asymmetries. We will also have biases to correct, but ours will differ from theirs in important ways ({\em CP}-symmetric initial state, no $B$ background, much lower charged multiplicities). It will be crucial to have independent corroborating evidence for these subtle measurements, such as we and LHC$b$ can provide.

\section{Proposed apparatus}

\begin{figure}[t]
\includegraphics[width=.7\linewidth,trim=275 172 100 65,clip
]{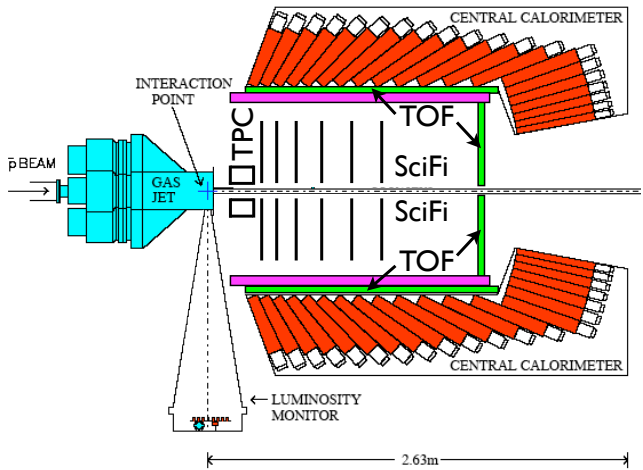}
\vspace{-.25in}
\caption[Sketch of proposed TAPAS apparatus]{Sketch of proposed TAPAS apparatus: a 1\,T solenoid surrounds a small, high-rate Time Projection Chamber and fine-pitch  scintillating-fiber detectors, 
and is surrounded by precision TOF counters, all within the existing E760/835 Central Calorimeter.  
A return yoke (not shown) is needed for proper functioning of calorimeter phototubes.}\label{fig:spect}
\end{figure}

The medium-energy antiproton-annihilation studies described above can all be carried out with a common apparatus, which can be assembled relatively quickly and cost-effectively thanks to the availability of key existing components: a barrel electromagnetic lead-glass calorimeter from E760/835~\cite{CalNIM}, a thin superconducting solenoid from BESS~\cite{BESS}, the D\O\ scintillating-fiber readout system~\cite{SciFi}, and plentiful trigger and data-acquisition electronics from D\O\ and CDF. Augmented with a small, high-rate TPC, new, thin, fine-pitch scintillating-fiber planes, and picosecond time-of-flight detectors currently under development~\cite{psec}, these can form a very powerful general-purpose spectrometer (Fig.~\ref{fig:spect}) for the low-multiplicity hadronic events that are produced by ${\overline p}p$ or ${\overline p}N$ annihilation in this energy range.

%

\subsection{Targets}

While the previous Fermilab Antiproton Source experiments E760 and E835 used internal gas-jet targets, a simpler approach is to use solid targets in the Antiproton Accumulator beam halo \`a la HERA-B~\cite{HERA-B}.  
Thin metal wire 
targets can serve to localize the production vertex, thereby facilitating charm-event identification via decay-vertex reconstruction. However, some of the proposed physics topics benefit from the elimination of Fermi motion and minimal beam 
energy loss provided by a hydrogen target. In particular, scanning of the beam energy across the resonance for precision determination of charmonium and $X(3872)$ masses and widths requires a hydrogen target. This can be accommodated by use of a windowless frozen target~\cite{Ishimoto-private}, based on successful experience with previous designs~\cite{Ishimoto}.

Further details of the proposed TAPAS experiment may be found in the proposal~\cite{TAPAS-proposal}.

%

\section{Antihydrogen measurements}

Antihydrogen ($\overline{\rm H}$) is produced whenever relativistic antiprotons pass through matter, via $e^+e^-$ pair production in the field of the nucleus. Fermilab E862 detected 99  $\overline{\rm H}$ atoms, produced in the E835 hydrogen cluster-jet target, with zero background, and confirmed the theoretical production rate~\cite{Blanford1}. A method to measure energy levels of $\overline{\rm H}$ in flight has been devised~\cite{Blanford2}. This is a measurement that can be performed parasitically whenever there are antiprotons in the Accumulator. In preparation for such a program, a thin, gold-plated carbon foil was recently installed in the Accumulator in order to verify the expected production rate and assess the feasibility of such a program.

Measuring the gravitational acceleration of antimatter in the field of the Earth is a unique test of the weak equivalence principle of General Relativity~\cite{Fischler-Lykken-Roberts}. The rate of fall of slow $\overline{\rm H}$ atoms can be measured quite precisely using an atom interferometer, with precision $\Delta g/g\sim10^{-4}$ using matter gratings and $\sim10^{-9}$ with laser techniques, and methods to decelerate and trap antiprotons at Fermilab in order to form the needed slow-$\overline{\rm H}$ beam have been described~\cite{P981}.

\section{Outlook}

Reconfiguration of the Debuncher and Accumulator rings of the Antiproton Source has been proposed in order to form the muon and proton beams respectively needed for the $g-2$ and Mu2e experiments at Fermilab. The $g-2$ configuration is potentially compatible with antiproton running, which requires the Accumulator all the time but the Debuncher only 10 to 20\% of the time (i.e., during antiproton stacking), while $g-2$ requires the Debuncher all the time (as a $\pi$-to-$\mu$ decay channel) but not the Accumulator. The planned Mu2e configuration is incompatible with antiproton running; however, Mu2e's likely 2018 start leaves a several-year antiproton window of opportunity.

While the TAPAS proposal has yet to obtain approval at Fermilab, the collaboration and proposal are being strengthened in order to enhance the prospects for such approval. It is hoped that apparatus assembly and development of the needed software and firmware can commence not long after the anticipated end of the Tevatron run in Sept., 2011. Data-taking could begin by 2014. Further development of the AGE proposal awaits demonstration (in the Raizen lab at UT Austin) of the beam-stopping technique needed for $10^{-9}$ resolution.



\begin{thebibliography}{99}
%
%
\bibitem{Pasquinelli-etal}
R. Pasquinelli {\it et al.}, ``Progress in Antiproton Production at the Fermilab Tevatron Collider," PAC09, paper TU6PFP075 (2009).

\bibitem{Tandean-Valencia-PRD67}
J. Tandean, G. Valencia, Phys.\ Rev.\ D {\bf 67}, 056001 (2003).

\bibitem{Tandean}
J. Tandean, Phys.\ Rev.\ D {\bf 70}, 076005 (2004).

\bibitem{Tandean-Valencia}
J. Tandean, G. Valencia, Phys.\ Lett.\ B {\bf 451}, 382 (1999).

\bibitem{Chang-etal}
D. Chang, X.-G. He, and S. Pakvasa, Phys.\ Rev.\ Lett.\ {\bf 74}, 3927 (1995).

\bibitem{He-etal-SUSY}
X.-G. He, H. Murayama, S. Pakvasa, G. Valencia, Phys.\ Rev.\ D {\bf61}, 071701(R) (2000).

\bibitem{Tandean-private}
J. Tandean, private communication.

\bibitem{website}
``New Pbar Experiments at Fermilab" web page, {\tt http://capp.iit.edu/hep/pbar/}.

\bibitem{E871}
R.\,A. Burnstein {\it et al.}\  [HyperCP Collaboration], 
Nucl.\ Instrum.\ Meth.\ A {\bf 541}, 516 (2005).

\bibitem{Johansson}
T. Johansson, in {\sl Proc.\ Eighth Int.\ Conf.\ on Low Energy Antiproton Physics (LEAP '05)}, Bonn, Germany, 16--22 May 2005, AIP Conf.\ Proc.\ {\bf 796}, 95 (2005).

\bibitem{Baltay}
C. Baltay {\it et al.}, Phys.\ Rev.\ {\bf 140}, B1027 (1965).

\bibitem{HERAG}
High-Energy Reactions Analysis Group, report CERN-HERA-84-01 (1984).

\bibitem{Materniak}
C. Materniak, presented at {\sl Eighth Int.\ Conf.\ on Hyperons, Charm and Beauty Hadrons (BEACH08)}, Columbia, SC,  22--28 June 2008,  {\tt http://beach2008.sc.edu/includes/\\ documents/sessions/materniak.talk.pdf}; 
C. Materniak {\it et al.}, to be submitted to Phys.\ Rev.\ Lett.

\bibitem{Parketal}
H.\,K. Park {\it et al.}\  [HyperCP Collaboration], Phys.\ Rev.\ Lett.\ {\bf 94}, 021801 (2005).

\bibitem{P0models}
See e.g.\ X.-G. He, J. Tandean, G. Valencia, Phys.\ Lett.\
B {\bf 631} (2005) 100; 
 N.\,G. Deshpande, G. Eilam, J. Jiang, Phys.\
Lett.\ B {\bf 632} (2006) 212.

\bibitem{SuperLEAR}
N. Hamann {\it et al.}\  [CP-Hyperon Study Group], report CERN/SPSLC 92019, SPSLC/M491, 30 March 1992.

\bibitem{McGinnis-Stancari-Werkema}
D.\,P. McGinnis, G. Stancari, S.\,J. Werkema, Nucl.\ Instrum.\ Meth.\ A {\bf 506}, 205 (2003).

\bibitem{Braaten-Stapleton}
E. Braaten and J. Stapleton, Phys.\ Rev.\ D {\bf 81}, 014019 (2010).

\bibitem{E760-chi}
T.\,A. Armstrong  {\it et al.}\  [E760 Collaboration], Phys.\ Rev.\ D {\bf 47}, 772 (1993).

\bibitem{E835-psi-prime}
M. Andreotti {\it et al.}\  [E835 Collaboration], Phys.\ Lett.\ B {\bf 654} (2007) 74.

\bibitem{ELQ}
E. Eichten, K. Lane, and C. Quigg, Phys.\ Rev.\ D {\bf 73}, 014014 (2006); Erratum-{\it ibid.} 
{\bf 73}, 079903 (2006). 

\bibitem{QWG-Yellow}
N. Brambilla {\it et al.}\  [Quarkonium Working Group], {\sl Heavy Quarkonium Physics}, CERN Yellow Report CERN-2005-005 (2005).

\bibitem{molecule}
N.\,A. T{\o}rnqvist, Phys.\ Lett.\ B {\bf 590} (2004) 209.

\bibitem{Ishimoto-private}
S. Ishimoto, private communication.

\bibitem{Ishimoto} 
S. Ishimoto {\it et al.}, 
Nucl.\ Instrum.\ Meth.\ A {\bf 480}, 304 (2002); 
S. Ishimoto {\it et al.}, ``Thin Windowless Solid Hydrogen Target," RCNP-Prog-Report-426, Research Center for Nuclear Physics, Osaka Univ.

\bibitem{ditto}
E. Braaten, Phys.\ Rev.\ D {\bf 73},  011501R (2006).


\bibitem{Braaten-X-3872}
E. Braaten, Phys.\ Rev.\ D {\bf 77}, 034019 (2008).

\bibitem{HFAG}
Heavy-Flavor Averaging Group, {\tt http://www.slac.stanford.edu/xorg/hfag/}.

\bibitem{ICHEP2010}
M.\,E. Mattson, ``Precision measurements of CP violation and $D^0$--${\overline D}{}^0$ mixing at CDF," presented at {\sl 35th Int.\ Conf.\ on High Energy Physics}, Paris, France, July 22--28, 2010,  {\tt http://indico.cern.ch/contributionDisplay.py?contribId=1082\&confId=73513}.

\bibitem{Bigi-Uraltsev}
See e.g.\ I.\,I. Bigi and N. Uraltsev, Nucl.\ Phys.\  B {\bf 592}, 92 (2001).

\bibitem{Bigi09}
I.\,I. Bigi, ``No Pain, No Gain -- On the Challenges and Promises of Charm Studies," presented at Charm09, Leimen, Germany, May 2009,  
arXiv:0907.2950 [hep-ph].

\bibitem{Grossman-etal}
See e.g.\ Y. Grossman, A.\,L. Kagan, Y. Nir, Phys.\ Rev.\ D {\bf 75}, 036008 (2007).

\bibitem{PANDA-LoI}
M.\ Kotulla {\em et al.}\  [{$\overline{\rm P}$}ANDA Collaboration], Letter of Intent (2004).

\bibitem{Quigg-Eichten}
E. Eichten and C. Quigg, private communication.

\bibitem{Braaten-charm-est}
Using Eq.\,5 of Ref.\,\protect\cite{Braaten-X-3872}, we obtain 1.3\,$\mu$b.

\bibitem{Titov-Kampfer}
A.\,I. Titov and B. K\"ampfer, Phys.\ Rev.\ C {\bf 78}, 025201 (2008).

\bibitem{Titov}
A. Titov, private communication.

\bibitem{A-dep}
M.\,J. Leitch {\it et al.}, Phys.\ Rev.\ Lett.\ {\bf 72}, 2542 (1994).


\bibitem{Lourenco}
C. Louren\c{c}o, H.\,K. W\"ohri, Phys.\ Rep.\ {\bf 433} (2006) 127--180.


\bibitem{PDG}
K. Nakamura {\it et al.}\   [Particle Data Group], J. Phys.\ G {\bf 37}, 075021 (2010).

\bibitem{Petrov}
See e.g.\ A.\,A. Petrov, arXiv:0806.2498 [hep-ph], and references therein.

\bibitem{Buchalla-etal}
See e.g.\ Sec.\ 3.9 of M. Artuso, G. Buchalla, {\it et al.}, Eur.\ Phys.\ J. C {\bf 57}, 309--492 (2008). 

\bibitem{CalNIM}
L.\ Bartoszek {\it et al.}\  [E760 Collaboration], Nucl.\ Instrum.\ Meth.\ A {\bf 301}, 47 (1991).

\bibitem{BESS}
Y. Makida {\it et al.}, Adv.\ Cryo.\ Eng.\ {\bf 37A}, 167 (1992); 
Y. Makida {\it et al.}, IEEE Trans.\ Appl.\ Supercond.\ {\bf 5}, 174 (1995).


\bibitem{SciFi}
M. Ellis {\it et al.}, ``The design, construction, and performance of the MICE scintillating fibre trackers," Nucl.\ Instrum.\ Meth.\ A, in press, doi:10.1016/j.nima.2011.04.041; 
A. Bross {\it et al.},  Nucl.\ Instrum.\ Meth.\ A {\bf 477},  172 (2002).

\bibitem{psec}
H. Frisch, Univ.\ of Chicago, private communication.

\bibitem{HERA-B}
K. Ehret, Nucl.\ Instrum.\ Meth.\ A {\bf 446}, 190 (2000).

\bibitem{TAPAS-proposal}
L. Bartoszek {\it et al.} [TAPAS Collaboration\footnote{TAPAS and AGE Collaborations: Arizona: A. Cronin; 
Bartoszek Eng.: L.~Bartoszek; 
CEA Saclay: P.~Colas, I.~Giomataris; Duke: T\,J.~Phillips; 
Fermilab: G.~Apollinari, D.\,R.~Broemmelsiek, C.\,N.~Brown, D.\,C.~Christian, 
P.~Derwent, K.~Gollwitzer, A.~Hahn, P.~Kasper, J.~Lewis, V.~Papadimitriou, 
G.~Stancari, M.~Stancari, R.~Stefanski, J.\,T.~Volk,
 S.~Werkema, W.~Wester,
 H.\,B.~White, G.\,P.~Yeh; 
Ferrara: W.~Baldini;
First Point: R.\,G.~Greaves;
Genoa: M.~Macri, M.~Marinelli;
Hbar Tech.: G.\,P.~Jackson;
Houston: K.~Lau;
IIT Chicago: P.\,W.~Johnson, D.\,M.~Kaplan, T.\,J.~Roberts, J.~Terry, Y.~Torun, C.\,G.~White;
IIT Hyderabad: A.~Giri;
ITEP Moscow: A.~Drutskoy;
KSU: G.\,A.~Horton-Smith, B.~Ratra;
Korea: B.\,R.~Ko;
KyungPook: H.\,K.~Park;
Lebedev Inst.: O.~Piskunova;
Longwood: T.~Holmstrom;
Luther Coll.: T.\,K.~Pedlar;
Michigan: H.\,,R.~Gustafson;
Moscow State: M.~Merkin;
NASA MSFC: J.\,B.~Pearson;
NIU: M.~Fortner;
Northwestern: J.~Rosen;
Notre Dame: M.~Wayne;
Oxford: F.~Azfar;
PNNL: D.\,M.~Asner;
SMU: T.~Coan;
St.\ Xavier: A.~Chakravorty;
Texas: M.\,G.~Raizen;
Uppsala: T.~Johansson
 }], ``Proposal 986: Medium-Energy Antiproton Physics with The Antiproton Annihilation Spectrometer (TApAS) at Fermilab,"\\ {\tt http://www.fnal.gov/directorate/program\_planning/Nov2010PACPublic/\\ 986\_TApAS\_Proposal\_2010\_11\_10.pdf}; see also \protect\cite{website}.
\addtocounter{footnote}{-1}
\bibitem{Blanford1}
G. Blanford {\it et al.}, Phys.\ Rev.\ Lett.\ {\bf 80}, 3037 (1998).

\bibitem{Blanford2}
G. Blanford {\it et al.}, Phys.\ Rev.\ D {\bf 57}, 6649 (1998).

\bibitem{Fischler-Lykken-Roberts}
M. Fischler, J. Lykken, T. Roberts,
``Direct observation limits on antimatter gravitation,"
FERMILAB-FN-0822-CD-T, arXiv:0808.3929 [hep-th] (2008); 
M.\,M. Nieto, J.T. Goldman,
``The Arguments against `antigravity' and the gravitational acceleration of antimatter,"
Phys.\ Rept.\ {\bf 205},  221 (1991).

\bibitem{P981}
A.\,D.\ Cronin {\it et al.} [AGE Collaboration\footnotemark], ``Letter of Intent: Antimatter Gravity Experiment at Fermilab," February 3, 2009.
\end{thebibliography}


\end{document}